\documentclass[12pt]{article}
\pdfoutput=1
\usepackage{amsmath,amsfonts,graphicx,color,bbm,tikz,bm,setspace}
\usepackage{comment}
\usepackage[nosort]{cite}
\usepackage{subfigure}
\usepackage{qcircuit}
\usepackage{multirow}
\usetikzlibrary{calc,positioning}
\usetikzlibrary{patterns,arrows,decorations.pathreplacing}
\usepackage{caption}
\usepackage{ulem}
\tikzset{>=stealth}
\usepackage{lipsum}
\usepackage{tabularx}
\usepackage{fullpage}
\usepackage{hyperref}
\usepackage{bbold}
\usepackage[english]{babel}

\usepackage{listings}
\usepackage{tcolorbox}

\newcommand{\mathsym}[1]{{}}
\newcommand{\unicode}[1]{{}}

\makeatletter\@addtoreset{equation}{section}\makeatother

\newcommand{\be}{\begin{equation}}
\newcommand{\ee}{\end{equation}}
\def\beq{\begin{equation}}
\def\eeq{\end{equation}}
\newcommand{\bea}{\begin{eqnarray}}
\newcommand{\eea}{\end{eqnarray}}
\newcommand{\Tr}{{\rm Tr\,}}

\newcommand{\ket}[1]{{\left| {#1} \right>}}

\renewcommand{\title}[1]{\vbox{\center\LARGE{#1}}\vspace{3mm}}
\renewcommand{\author}[1]{\vbox{\center{#1}}\vspace{3mm}}

\newcommand{\email}[1]{\vbox{\center\tt#1}\vspace{3mm}}

\begin{document}

\begin{center}
{\large {\bf Asymptotic Quantization of Palatini Action }}

\author{A.P.Balachandran$^a$}

{$^a${\it Physics Department, Syracuse University, NY 13244-1130, USA }}
\vskip0.1cm
\vskip0.1cm

\email{apbal1938@gmail.com}

\vskip 0.5cm 

\end{center}


\abstract{
\noindent 
The Palatini action is based on  vector valued one forms or frames and  $SL(2,\mathbb{C})$ connections on $\mathbb{R}^4$. Using the spacetime split  $\mathbb{R}^4  = \mathbb{R}^3\oplus\mathbb{R}^1$, the Gauss law in this paper is treated  on  a {\it Hilbert space}. This is achieved by noticing that quantum operators act on a {\it complex Hilbert space}  and $SL(2,\mathbb{C})$ is just the complexification of  the compact $SU(2)$
in the self-dual $\left(1/2,0\right)$  representation
used for the Ashtekar variables. This leads to a treatment of small and large gauge transformations and superselection sectors. An explicit representation of theta vacua and its attendant 'spin-isospin mixing'  are also shown. It is argued that the Gauss law algebra replaces that of diffeomorphisms in the Palatini approach: operators implementing the latter with the correct algebraic relations do not seem  available. (Those obtained by multiplying Gauss law operators with fields do not  have the correct commutators.)
}


\section{Introduction}
\label{sec:Introduction}
The spacetime in this paper is $\mathbb{R}^4$ with fixed initial value slice $\mathbb{R}^3$.

We work with Palatini action. This is the model where the Lorentz group is gauged and has the action 
\begin{eqnarray}\label{eq:palatiniAction}
    S & = & \frac{1}{4G}\int~\epsilon_{abcd}e^a\wedge e^b\wedge F^{cd}, \nonumber \\
    F & = & dA + A\wedge A.
\end{eqnarray}
$A$ is the $SO(3,1)$ connection, $A=\left(A^{cd}\right)$, $A^{cd}$ a real-valued one-form $A^{cd}_\mu dx^\mu$. Its indices $\mu$ transform as four-vectors. The $\epsilon$ is the Levi-Civita symbol with $\epsilon_{0123}=+1$, $G$ the gravitational constant. The frames $e$ are fields which at each $x$ are real-valued one-forms with values in a vector space $V$ on which $SO(3,1)$ acts by the four-vector representation. It has the $SO(3,1)$-invariant metric with mostly $+$ signature. $V$ is not a Hilbert space.

Classically, if the frame field is invertible at every $x$, $S$ can be reduced to the Einstein-Hilbert action. But not otherwise. The latter case leads to new physics as shown by Kaul and Sengupta \cite{Kaul_2016,Sengupta_2023}.

In quantum theory, the property of fields are determined by the quantum states and their representation through the $GNS$ construction. `Invertibility' of a field, an operator-valued distribution, has also to be given a meaning. Such issues will be addressed later. 

We do not need the details of the canonical quantisation of $S$. But as in $QCD$, we note that if $A^{ab}_i$ $\left(i=1,2,3\right)$ is the connection on the initial spatial hypersurface and $E^{cd}_i$ its conjugate electric field, we get the `formal' Gauss law
\begin{eqnarray}\label{eq:1.2}
    D.E\ket{.} =0
\end{eqnarray}
on allowed quantum states. (The indices are raised and lowered with the flat 
Lorentzian metric with mostly plus signature.)

As \eqref{eq:palatiniAction} is already in the $p~dq$ form in time derivatives, as seen from the term 
\begin{eqnarray}
    \frac{1}{2G}\epsilon_{abcd}\epsilon^{ij0k}e^a_ie^b_jF^{cd}_{0k}
\end{eqnarray}
got from varying $A_0$, where $F_{0k}$ has $\partial_0A^k$, the conjugate field $E_i$ is seen to be 
\begin{eqnarray}
    E^i_{cd} = \epsilon^{ijk}\epsilon_{abcd}e^a_je^b_k.
\end{eqnarray}

\section{Introducing Self-Dual Connections}
\label{sec:IntroSelfDualConnections}
Since the fields act also on an $SO(3,1)$ representation, the vector state $\ket{.}$ has to be of the form
\begin{eqnarray}
    \ket{0}\otimes V
\end{eqnarray}
where fields with fixed indices act on $\ket{0}$ (after smearing with test functions).

But $V$ is not a Hilbert space while quantum theory prefers that. We need an approach to resolve this problem.

Following Ashtekar \cite{AshtekarPRL,AshtekarPRD}, Jacobson and Smolin \cite{Jacobson_1988}, and Samuel \cite{Samuel1987}, $V$ is replaced by $\mathbb{C}^2$ carrying the representation $\left(1/2,0\right)$ of $SL(2,C)$. The basis for the $SL(2,C)$ Lie algebra then is 
\begin{eqnarray}
    \sigma_{ab},~~~a)~\sigma_{ij} & = & \frac{1}{2}\epsilon_{ijk}\tau_k,~~~b)~\sigma_{0i} = \frac{i}{2}\tau_i \\
    \tau_i & = & \textrm{Pauli matrices}. \nonumber 
\end{eqnarray}
They transform under $SL(2,C)$ by the adjoint representation.

For the $\left(0,1/2\right)$ representation, a) and b) are replaced by their complex conjugates. In these representations, the standard positive definite metric on $\mathbb{C}^2$ is $SU(2)$ invariant and $SL(2,C)$ is just the complexification of $SU(2)$.

In the $\left(1/2,0\right)$ representation, we can write an $SU(2)$ connection and its conugate field (suppressing the `spatial' index) as
\begin{eqnarray}
    A = A^i\tau_i~~,~~E-=E_i\tau_i
\end{eqnarray}
where $A^i$, $E_i$ are 3 independent pairwise conjugate self-adjoint fields acting on a complex Hilbert space. They can be quantised as such.

Now if $g$ is in $\left(1/2,0\right)$, we can write 
\begin{eqnarray}
    g = h~\rm{exp}(\Delta)
\end{eqnarray}
where $h$ is unitary and is in $SU(2)$ while $\rm{exp}(\Delta)$ is self-adjoint and of determinant 1 :
\begin{eqnarray}
    \rm{exp}(\Delta) = \rm{exp}(\tau_i\theta_i), ~~~\theta_i\in\mathbb{R}.
\end{eqnarray}
Its adjoint action on $E$ for example is 
\begin{eqnarray}
    E_i\tau_i \rightarrow E_i,~~~~\rm{exp}(\Delta)\tau_i\rm{exp}(-\Delta)=\tau_i'.
\end{eqnarray}
$\tau_i$'s are traceless and fulfill
\begin{eqnarray}
    \Tr \tau_i'\tau_j' = 2\delta_{ij},
\end{eqnarray}
but they are not self-adjoint. Hence
\begin{eqnarray}
    \tau_i' = R(C)_{ij}\tau_j
\end{eqnarray}
where $R(C)$ is an element of the complex orthogonal group $SO(3,C)$:
\begin{eqnarray}
    R(C)\in SO(3,C).
\end{eqnarray}
Hence under $\textrm{exp}(\Delta)$, $E_i$, becomes $E'_i$ where
\begin{eqnarray}
    E_i'=E_jR(C)_{ji}
\end{eqnarray}
which is a well-defined operator on the complex Hilbert space.

\section{Treatment of the Gauss Law}
\label{sec:GaussLaw}
As in $QCD$, and for reasons explained before \cite{Balachandran_2019}, we write \eqref{eq:1.2} using the compactly supported test functions $\chi^\alpha$ as 
\begin{eqnarray}
    Q(\chi)~\ket{.} & = & 0, \label{eq:3.1}\\
    Q(\chi) & = & \int \Tr\left[\left(D_i\chi \right)_i E^i\right]
\end{eqnarray}
where $\chi=\chi^\alpha\lambda_\alpha$, with $\lambda_\alpha$ a basis for the appropriately chosen Lie algebra. The trace is over Lie algebra indices.

One requires \eqref{eq:3.1} for every such $\chi$, even for complex (compactly supported) functions  $\chi^\alpha$.

$\lambda^\alpha$ here is that of $\left(1/2,0\right)$ representation. So $\chi^\alpha\lambda_\alpha=\chi_1^\alpha\tau_\alpha + \chi_2^\alpha(i\tau_\alpha)$ where $\chi^\alpha$ are complex.

Note that if \eqref{eq:3.1} is the case for $\chi^\alpha$ real, $cQ(\chi) = Q(c\chi)$ also vanishes on $\ket{}$ for any complex $c$ and that covers $SL(2,C)$ where $\chi$'s are complex. As real $\chi^\alpha$ is a special case of complex $\chi^\alpha$, converse is also valid.

All observables $K$ (see later) fulfill
\begin{eqnarray}\label{eq:3.3}
    \left[ K, Q(\chi)\right]\ket{.} =0
\end{eqnarray}
for every such $\chi$.

It is enough if \eqref{eq:3.3} is fulfilled for real $\chi^\alpha$ by the previous remarks since \eqref{eq:3.3} for real $\chi^\alpha$ implies \eqref{eq:3.3} also for complex $\chi^\alpha$.

$Q(\chi)$ generate `small' gauge transformations. They can be extended to generators $Q(\xi)$ of `large' gauge transformations on using test functions $\xi$ which do not necessarily vanish at infinity. By locality, $Q(\xi)$ still fulfill
\begin{eqnarray}\label{eq:3.4}
    \left[ K, Q(\xi)\right]\ket{.} =0
\end{eqnarray}
but $Q(\xi)$ need not vanish on allowed quantum states.

Also it is the case that 
\begin{eqnarray}\label{eq:3.5}
    \left[Q(\xi), Q(\xi')\right] = Q(\left[\xi, \xi'\right])
\end{eqnarray}
and hence also that 
\begin{eqnarray}\label{eq:3.6}
    \left[Q(\xi), Q(\chi)\right] = Q(\left[\xi, \chi\right]).
\end{eqnarray}
\eqref{eq:3.6} generates a small gauge transformation, $\left[\xi, \chi \right]$ being compactly supported.

$Q(\xi)$ generate the Lie algebra of all gauge transformations $\cal{A}$ and $Q(\chi)$ generate the Lie algebra of `small' gauge transformations $\cal{A}_{\rm{small}}$. As the commutator of two such $\xi$ may be a $\chi$, large gauge transformations do not close under commutation.

In gauge theories like $QED$ and $QCD$, the spectra of generators of large gauge transformations label the representations of the superselection sectors.

Now because $A_{i,m}$ and $E^{j,n}$ fulfill ($m$ and $n$ are form and vector field indices) 
\begin{eqnarray}
    \left[A_{i,m}(x), E^{j,n}(y)\right] = i\delta(x,y)\delta_{i,j}\delta_m^n,
\end{eqnarray}
they have a Hilbert space representation : for each $i$, they create a Fock space on $\ket{0}$ for $A_i$, $E_i$.

The indices $m$ and $n$ are of $\mathbb{R}^3$ and can be incorporated without indefinite metric. Using fields localised on strings of $\mathbb{R}^3$ \cite{Mund_2020}, one can construct gauge invariant operators for them as discussed below.

\subsection{String Localised Fields}
\label{subsec:stringLFields}
The tensorial fields $E$ or $dA+A\wedge A$ do not fulfill the Gauss law condition
\begin{eqnarray}
    Q(\chi)\ket{\cdot} = 0.
\end{eqnarray}
This issue can be treated below using the following property of the Dirac-Wilson line following \cite{Mund_2020}.

In brief, the treatment is as follows. For a spacelike vector $e$ with
\begin{eqnarray}\label{eq:3.9}
    e.e = -1,~~~~\rm{metric~mostly~minus,}
\end{eqnarray}
consider 
\begin{eqnarray}
    W(x,e) = P\textrm{exp}\int d\tau e^\mu A_\mu(x+\tau e)
\end{eqnarray}
$\mu$ of $A_\mu$ being a form index and $A_\mu = A^i_\mu\tau_i$.

Under the action of $g$ in the $\left(1/2,0\right)$ representation,
\begin{eqnarray}\label{eq:3.11}
    W(x,e) \rightarrow g(e_\infty)W(x,e)g^{-1}(x)
\end{eqnarray}
where 
\begin{eqnarray}
    e_{\infty} := \rm{lim}~\tau e.~ \textrm{as}~\tau\rightarrow\infty.
\end{eqnarray}
Hence a field such as a spinor $\psi=$ ($\psi$ transforming by $\left(1/2,0\right)$ : 
\begin{eqnarray}
    \psi \rightarrow g\psi
\end{eqnarray}
can be made immune to small gauge transformations by converting it to the string-localised field 
\begin{eqnarray}
    W(x,e)\psi.
\end{eqnarray}
But under a large gauge transformation, it transforms at $\infty$ as in \eqref{eq:3.11}:
\begin{eqnarray}
    W(x,e)\psi \rightarrow g(e_\infty)W(x,e)\psi.
\end{eqnarray}

\section{Theta Vacua for Palatini}
\label{sec:thetaVP}
These generate superselection sectors in $QCD$ and are present in Einstein-Hilbert gravity as discussed in \cite{AshtekarCP}. They are present in the Palatini approach too as shown below.

The presentation follows \cite{balachandran2023spin12gluons} and has unusual properties.

Consider 
\begin{eqnarray}\label{eq:4.1}
    \textrm{exp}(i\frac{\theta}{4}K(A))\ket{0},~~K(A) = \frac{1}{8\pi^2}\Tr{\left(A\wedge dA+\frac{2}{3}A\wedge A\wedge A\right)}
\end{eqnarray}
where we assume as usual that there is a vacuum $\ket{0}$ fulfilling all the constraints and $K(A)$ is the $\left(1/2,0\right)$ Chern-Simons term. The trace is over the $\left(1/2,0\right)$ representation. $\ket{0}$ is thus an $SL(2,C)$-singlet.

Let 
\begin{eqnarray}
    U(x) & = & \cos{\theta(r)}+i\tau.\hat{x}\sin{\theta(r)} \nonumber \\
     & = & \textrm{exp}(i\theta(r)\tau.\hat{x}), \nonumber \\
     \theta(0) & = & \pi,~\theta(\infty)=0
\end{eqnarray}
be a winding number one chiral soliton of Skyrme model \cite{BalBook}. From it, we pick the test function 
\begin{eqnarray}
    \xi_{\textrm{skyrme}} = \theta(r)\tau.\hat{x}
\end{eqnarray}
and the generator of a large gauge transformation
\begin{eqnarray}\label{eq:4.4}
    Q(\xi_{\textrm{skyrme}}).
\end{eqnarray}
Then, $\rm{exp}(iQ(\xi_{\rm{skyrme}}))$, being a winding number 1 gauge transformation, has eigenvalue $\rm{exp}(i\theta)$ for the vector \eqref{eq:4.1}.

As observed in the paper on gluon theta vacuum \cite{balachandran2023spin12gluons}, rotation generators of orbital angular momentum do not commute with $Q(\xi_{\rm{skyrme}})$ and hence are spontaneously broken. They have to be renormalised from $L_i$ to 
\begin{eqnarray}
    J_i = L_i + Q\left(\frac{\tau_i}{2}I\right)
\end{eqnarray}
where $I$ is the constant function with value 1. We must also change \eqref{eq:4.1} to 
\begin{eqnarray}\label{eq:4.6}
    \textrm{exp}(i\frac{\theta}{4}K(A))\ket{0}\otimes\mathbb{C}^2
\end{eqnarray}
where $\mathbb{C}^2$ carries the action of $Q\left(\frac{\tau_i}{2}I\right)$. This is the phenomenon of `mixing spin with isospin' \cite{Jackiw1976,Hasenfratz1976} (See also Sec. 5 of \cite{balachandran2006lecturesfuzzyfuzzysusy}).

Also since the $2\pi$ rotation is
\begin{eqnarray}
    \textrm{exp}(i2\pi J_i) = -1,
\end{eqnarray}
the vector states from the folium of \eqref{eq:4.6} are spinorial as in Friedman and Sorkin's `Spin 1/2 from Gravity' \cite{Friedman1980}.

As observed in the paper on the gluon theta vacua, there is an identification of elements of spatial $\mathbb{R}^3$ and $\left(1/2,0\right)$ rotations in \eqref{eq:4.4}. If $\tau_i$ is replaced by $\tau_i'=s\tau_i s^{-1}$ where $s\in SU(2)$, the winding number will not change, but
\begin{eqnarray}
    Q\left(\xi'_{\rm{skyrme}} \right) - Q\left(\xi_{\rm{skyrme}} \right) = Q\left(\xi'_{\rm{skyrme}}-\xi_{\rm{skyrme}} \right)
\end{eqnarray}
is a large gauge transformation so that the superselection sector is changed. So will the transformation of connection forms as $\tau_i$ are changed.

\section{On Local Observables in Gravity and Can They Affect the Line $x+\tau e$ ?}
\label{sec:localObservableG}
Already in $QFT$, it is impossible to make local measurements of operators like $\textrm{exp}(i\phi(\psi))$ where $\phi$'s are local fields and the test Schwartz functions $\psi$ are locally supported. This is already the case in quantum mechanics :  if $B(H)$ is its family of observables and $\omega$ a state, individual elements of $B(H)$ are defined only upto automorphisms $U$ of $B(H)$. If $U$ is an inner automorphism, there is no way to distinguish $a$ from $UaU^{-1}$ algebraically in $B(H)$. But $\omega(a)$ need not be equal to $\omega(UaU^{-1})$.

But if $U$ is an outer automorphism, it generally changes the superselection sector and the representation of observables, so it is reasonable to assume that observational difficulties come from inner automorphisms. 

Often, one gets around this problem by introducing an external frame. But is the external frame quantum ? It must be as everything is quantum. So we end up with an infinite regression which is being arrested by claiming that frames are classical. 

This is odd. Nature cannot be schizophrenic. Bohr, Hepp, Landsman and others get around this by showing that there is an emergent commutative algebra at `infinity' where classical measurement theory can be applied.

We have also addressed this problem and suggested a solution \cite{Balachandran_2022}.

So certainly, there are no exclusively local measurements possible in quantum gravity.

Now in gravity, it is usually assumed that spacetime is asymptotically flat and if $diff$ and $diff_0$ are large and small diffeos, $diff/diff_0$ is the Poincar\'{e} group. This lies at `infinity' and states presumably can detect them, being invariant under small diffeos by the gravitational Gauss law.

Remark : In the title, $e$ is in the direction of spacelike straight line as in \eqref{eq:3.9} defined by using a flat metric in $\mathbb{R}^3$.

We may thus prepare vector states invariant under the $SU(2)$ in the Sky group diagonalising $Q\left(X_3\frac{\tau_3}{2}\right)$ where $X_3$ is a function on $S^2$. As before, we get $C^\infty(S^2)\times C^\infty(S^2) $ from the two eigenvalues of $\frac{\tau_3}{2}$.

It is to be remarked that we quantise only $A_i$ and $E^j$ on the $\mathbb{R}^3$ hypersurface chosen as above where they are canonically conjugate. Or more generally we quantise $A_\mu e^\mu$ and its conjugate $E^\mu e_\mu$ where the chosen flat metric is used to raise and lower indices. As $e^\mu$'s are spacelike, issues of indefinite metric do not arise. 

The Lorentz group acts on the spacelike direction, and its unitary implementation runs into trouble as it has been seen in \cite{Mund_2020}.

Does the asymptotically ADM diffeo group also preserve the spacelike character of $e$? It does, as diffeos do not affect the signature of a metric. But again, can one implement diffeos by operators on a Hilbert space? The answer is unclear, seems no for diffeos which act at infinity on $e$. After all, Lorentz transformations which are part of the ADM group cannot be implemented if they change the superselection sector. So one can implement only rotations around the axis $e$. 

\section{Tentative Conclusions for Palatini Gravity}
\label{sec:conclusions}
\begin{enumerate}
    \item It does not seem equivalent to the Einstein-Hilbert gravity. In the $\left(1/2,0\right)$ formulation for gauge group, the $SU(2)$ connection on `complexification' gives the $SL(2,C)$ one. This complexification is automatic as the connections are operators on a complex Hilbert space when one quantises on an $\mathbb{R}^3$ slice. Diffeos change the direction of $e$, but it remains spacelike. So a Hilbert space quantisation seems possible much like in two-colour QCD.
    \item Generic diffeos which change $e$ to another $e'$ are spontaneously broken. 
    \item $QCD$ theta vacua can be adapted to quantum gravity. They can change their folia from bosonic to fermionic and conversely as in $QCD$.
    \item Where are the diffeo constraints ? The Palatini action contains no metric. If one {\it assumes} that the frame fields are invertible, then classically, the action can be reduced to the Einstein action.

    But one is engaged with {\it quantum theory} where the frames are operator-valued distributions. Their invertibility or otherwise should follow from the state defining the representation of the observables from the $GNS$ construction. Lacking this, the meaning of `invertibility' is not clear. 

    The substitution of `invertibility' with `unitarity' at each spacetime point $x$, $e*e(x)=\mathbb{1}$ on $V$ at each $x$, is equally problematic. The $*$ is not the star of the full Hilbert space, but just of $V$. But the product of quantum fields at the same $x$ is generally divergent.

    Can one at least write the generator of local diffeos using the Gauss law, multiplying it by some fields ? Such issues arise in the $ISO(2,1)$ Chern-Simons gravity \cite{WITTEN198846}. The suggested solution is to multiply the gauge constraint by a field and show that it generates local diffeos and vanishes on quantum states.

    But such operators do not fulfill commutation relations of the diffeo Lie algebra.

    The situation for the Palatini action seems similar. There may be multipliers of Gauss law constraints by fields which leads to apparent infinitesimal diffeos, but they are unlikely to have correct commutators. 
\end{enumerate}

\section*{Acknowledgments}
I have benefited form inputs from Babar Qureshi and Bruno Carneiro da Cunha. The help of Pramod Padmanabhan was invaluable in preparing this paper.

\bibliographystyle{acm}
\normalem
\bibliography{refs}

\end{document}